# Synchronization of Uncertain Fractional-Order Duffing-Holmes Chaotic System via Sliding Mode Control

S.H. Hosseinnia*, R. Ghaderi*,
A. Ranjbar N.*, S. Momani**

*Noushirvani University of Technology, Faculty of Electrical and Computer Engineering, P.O. Box 47135-484, Babol, Iran,
(a.ranjbar@nit.ac.ir) ,(h.hoseinnia@stu.nit.ac.ir)
** Department of Mathematics, Mutah University, P.O. Box: 7, Al-Karak, Jordan

Abstract: In this paper, a sliding mode controller is designed to synchronize a chaotic fractional-order system. To construct a corrective control input, a saturation function $sat(.)$, with a modified sliding surface is proposed. Finally, Chaos in the Duffing-Holmes system with fractional orders is investigated, and a numerical simulation (synchronizing fractional-order Duffing-Holmes _ Duffing-Holmes system) are presented to show the effectiveness of the proposed controller.

## 1. INTRODUCTION

Fractional calculus is an old mathematical topic from 17th century. Nowadays, it has been found that some fractional-order dynamic systems such as Chua circuit (Hartley *et. al.*, 1995), Duffing system (Arena *et. al.*, 1997, jerk model (Ahmad and Sprott, 2003), Chen system (Lu and Chen, 2006), Lü dynamic (Lu, 2006), Rossler model (Li and Chen, 2006), Arneodo system (Lu, 2005) and Newton–Leipnik model (sheu *et. al.*, 2006) can demonstrate chaotic behaviour. Over the past two decades, due to the pioneering work of Ott et al. (Ott *et. al., 1996)*, synchronization of chaotic systems has become more and more interesting for researchers in different area. The problem of designing a system, whose behaviour mimics that of another chaotic system, is called synchronization. Two chaotic systems are generally called drive (master) and response (slave) systems respectively. Recently, synchronization of chaotic fractional-order systems starts to attract increasing attention due to its potential applications in secure communication and control processing (Matignon, 1996). For example, in (Deng and Li, 2005a) chaos synchronization of two Lü systems has been studied. Synchronization of two chaotic fractional Chen and Chua systems have been presented in (Deng and Li, 2005b; Li *et. al.,* 2006), respectively. Variable structure control with sliding mode, which is commonly known as sliding mode control (SMC), is a nonlinear control strategy that is well known for its robust characteristics (Utkin, 1977). The main feature of SMC is that it can switch the control law very fast to drive the states of the system from any initial states onto a user-specified sliding surface, and to maintain the states on the surface for all subsequent time (Utkin, 1977; Phuah et. al., 2005). This controller achieves an insensitive to parametric uncertainties and external disturbances (Ertugrul and Kaynak, 2000; Slotine and Sastry, 1983) performance. In this paper, a SMC is presented to synchronize fractional-order Duffing-Holmes chaotic system.

This paper is organized as follows:
In second section, Sliding Mode Control is presented to synchronize the Fractional-order chaotic system in presence of uncertainty. Section 3 is devoted to apply the method on Fractional-order Duffing-Holmes chaotic system. The work will be simulated using $MATLAB^®$ 7.4 software. Finally, the result and conclusion will be concluded in section 4.

## 2. SLIDING MODE CONTROL OF FRACTIONAL ORDER SYSTEM

To design a sliding mode controller, a sliding surface must be primarily designed. This surface introduces the desired dynamic of system and therefore completes the movement the sliding control law along with the surface. Every outer state will be pulled into the surface in finite time. This controller is proposed to synchronize a chaotic system with fractional dynamic. The chaos synchronization problem means making two systems oscillate in a synchronized way. Let us call a particular dynamical system as master and a different dynamical system as a slave. The goal is to synchronize the slave with the master system. In order to achieve this synchronization, a nonlinear control system that obtains signals from the master system and controls the slave system should be designed. Let us consider master (equation 1) and slave (equation 2) systems with fractional order derivative as follows:

$$D^q x_1 = x_2 \qquad (1)$$
$$D^q x_2 = f(X,t)$$
$$D^q y_1 = y_2 \qquad (2)$$
$$D^q y_2 = f(Y,t) + \Delta f(Y,t) + d(t) + u(t)$$

where $X = [x_1, x_2]^T$ and $Y = [y_1, y_2]^T$ are the states of appropriate system in (1) and (2) respectively and $0 < q \leq 1$. Furthermore, $f(Y,t)$ is an unknown nonlinear function, $\Delta f(Y,t)$ is uncertainty and $d(t)$ is an acting disturbance against the performance of the system. The control input $u(t)$, is designed via sliding mode to perform a tracking task for the slave to follow the master. The sliding mode control will be designed according to the following procedure:

1- To construct a sliding surface that represents a desired system dynamics
2- To develop a switching control law such that sliding mode exists on every point of the sliding surface, and any states outside the surface are driven to reach the surface in a finite time.

To design the control law $u(t)$, the synchronization error is defined as:

$$e_i = x_i - y_i \qquad (3)$$

The sliding surface should be designed accordingly by:

$$S(t) = c_1 e_1 + c_2 e_2 \qquad (4)$$

where $c_1$ and $c_2$ will be chosen such that dynamic of the sliding surface will be vanished quickly. As soon as the state reaches to the surface, it stays there forever. This effect is usually called; the sliding mode is taken place. At this stage, the dynamic will be controlled by the dynamic of sliding mode. Therefore $c_1$ and $c_2$ must be designed in such a way the surface behaves a desired dynamic (Slotine, 1991). However, the sliding mode control will be deigned in two phases:

1. The reaching phase when $S(t) \neq 0$ and
2. The sliding phase by $S(t) = 0$.

A sufficient condition for the error to move from the first phase to the second is as follows:

$$S(t).\dot{S}(t) \leq 0 \qquad (5)$$

This condition is called the sliding condition. In the sliding phase $S(t) = 0$ and $\dot{S}(t) = 0$. The following equation converts the classic derivative into a fractional type.

$$\dot{S}(t) = D^{1-q}(D^q(S(t))) = 0 \qquad (6)$$
$$\text{if } D^q(S(t)) = 0 \rightarrow \dot{S}(t) = 0$$

Control signals in the following equation derive the dynamic to reach to the sliding surface:

$$D^q(S(t)) = c_1 D^q e_1 + c_2 D^q e_2 = \begin{bmatrix} c_1 & c_2 \end{bmatrix} \begin{bmatrix} D^q(x_1 - y_1) \\ D^q(x_2 - y_2) \end{bmatrix}$$
$$= \begin{bmatrix} c_1 & c_2 \end{bmatrix} \begin{bmatrix} e_2 \\ f(X) - f(Y) - \Delta f(Y,t) - d(t) - u_{eq}(t) \end{bmatrix} \qquad (7)$$
$$= \begin{bmatrix} c_1 & c_2 \end{bmatrix} \begin{bmatrix} e_2 \\ f(X) - f(Y) - \Delta f(Y,t) - d(t) \end{bmatrix} - c_2 u_{eq}(t)$$

Therefore, the equivalent control law is obtained as follows:

$$u_{eq}(t) = \begin{bmatrix} c_1/c_2 & 1 \end{bmatrix} \begin{bmatrix} e_2 \\ f(X) - f(Y) - \Delta f(Y,t) - d(t) \end{bmatrix}$$
$$= \frac{c_1}{c_2} e_2 + f(X) - f(Y) - \Delta f(Y,t) - d(t) \qquad (8)$$

In the reaching phase where $S(t) \neq 0$ and in order to satisfy the sliding condition in (5) another correction term $u_c(t)$, will be added to modify the control law (switching function). To investigate the stability, a proper Lyapunov function should be candidate to guarantee the stability. This will be defined as:

$$V_{SMC} = \frac{S^2(t)}{2} \qquad (9)$$

It should be noted that this function is a positive definite. An auxiliary goal is find a Lyapunov function with negative sign for the derivative. This will be satisfied if equation (10) is met.

$$D^q S(t) = -k_s.sat(S(t)) \rightarrow$$
$$\dot{S}(t) = -k_s.D^{1-q}(sat(S(t))) \qquad (10)$$

where $k_s$ is a positive constant gain. Meanwhile $sat(S(t))$ is defined as follows:

$$sat(S) = \begin{cases} +1 & S(t) > \varepsilon \\ t & -\varepsilon \leq S(t) \leq \varepsilon \\ -1 & S(t) < -\varepsilon \end{cases} \qquad (11)$$

where $\varepsilon$ is a small parameter. When (10) is substituted in the derivative of equation (9), these approaches to:

$$\dot{V}_{SMC}(t) = S(t).\dot{S}(t) = -k_s S(t).D^{1-q}(sat(S(t))) \quad (12)$$

It is obvious that equation (12) is negative definite. Therefore, the selection $k_s > 0$ guarantees the stability. The fractional derivative of equation (4) takes the following form:

$$D^q S(t) = c_1 e_2 + c_2[f(X) - f(Y) - \Delta f(Y,t) - d(t) - u(t)] \quad (13)$$

The control signal using both equations in (10) and (13) will be obtained as:

$$u(t) = u_{eq}(t) + u_c(t) = \frac{c_1}{c_2}e_2 + f(X) - f(Y) - \Delta f(Y,t) - d(t) + \frac{k_s}{c_2}sat(S(t)) \quad (14)$$

where,

$$u_c(t) = \frac{k_s}{c_2}sat(S(t)) \quad (15)$$

completes the control law. When $K_s$ is defined by $K_s = \frac{k_s}{c_2}$ the final control is obtained as follows:

$$u_c(t) = K_s sat(S(t)) \quad (16)$$

Where $K_s$ is the switching gain. In following section primarily, the chaos will be investigated in factional dynamic Duffing–Holmes. The sliding mode control will be applied to synchronize the system whilst it is perturbed with uncertainty and disturbance.

## 3. SYNCHRONIZATION OF UNCERTAIN FRACTIONAL ORDER DUFFING-HOLMES CHAOTIC SYSTEM

*3.1. System description*

In (Chang and Yan, 2005) an integer type of chaotic Duffing–Holmes is investigated. The chaos will be investigated when a fractional dynamic Duffing–Holmes. This system is considered as follows:

$$\begin{cases} D^q x_1 = x_2 \\ D^q x_2 = x_1 - \alpha x_2 - x_1^3 + \beta \cos(t) \end{cases} \quad (17)$$

Phase portrait of dynamic (17) for different values of $q$ is shown in Figure (1). Parameters $\alpha$ and $\beta$ are chosen 0.25 and 0.3, respectively. As it can be seen when $q$ is reduced the chaos is accordingly reduced and the behaviour approaches to an oscillatory dynamic. In order to synchronize the fractional system in (17), a sliding mode controller will be implemented. In this section, the fractional system in (17) will be used in a synchronization task as master. To do this, Duffing-Holmes system, which is perturbed by uncertainty and disturbance, will be considered as the slave by the following dynamic:

$$\begin{cases} D^q y_1 = y_2 \\ D^q y_2 = y_1 - \alpha y_2 - y_1^3 + \beta \cos(t) + \Delta f(Y,t) + d(t) + u(t) \end{cases} \quad (18)$$

where $\Delta f(Y,t) = 0.1\sin(t)\sqrt{y_1^2 + y_2^2}$ and $d(t) = 0.1\sin(t)$.

*3.2 Implementation*

A sliding mode controller is implemented to synchronize fractional Duffing-Holmes dynamic in MATLAB® 7.4 environment.

Initial conditions of the Master and Slave are considered as: $x_1(0) = 0.2$, $x_2(0) = 0.2$ and $y_1(0) = 0.1$ $y_2(0) = -0.2$, respectively. Regarding to equation (8) the equivalent control law will be of the following form:

$$u_{eq}(t) = \frac{c_1}{c_2}e_2 + e_1 - \alpha x_2 - x_1^3 + \alpha y_2 + y_1^3 - 0.1\sin(t)(\sqrt{y_1^2 + y_2^2} + 1) \quad (19)$$

Replacing equation (19) in (14) results the control law, which is as follows:

$$u(t) = \frac{c_1}{c_2}e_2 + e_1 - \alpha x_2 - x_1^3 + \alpha y_2 + y_1^3 - 0.1\sin(t)(\sqrt{y_1^2 + y_2^2} + 1) + K_s sat(S(t)) \quad (20)$$

The simulation result is shown in Figure 2 when parameters are chosen $c_1 = c_2 = 1$ and $K_s = 10$. The control signal, sliding surface and synchronization of states X and Y for $q = 0.98$ are shown in Figure (2, a) and (2, b). Those are also shown for different values of $q = 0.96$ and $q = 0.9$ in Figures (2.c, 2.d) and (2.e, 2.f) respectively. It should be noted that the control is triggered at $t = 20s$. The significance of the sliding mode control in 3 simulations for different values of $q$ is shown during the speed of synchronization of Master and Slave. This also verifies the robustness of the controller.

## 4. CONCLUSION

In this paper, chaos in the Duffing-Holmes system with fractional orders is studied whilst a sliding mode controller has proposed to synchronize uncertain chaotic fractional order systems. It has been shown that by a proper selection of the control parameters ($K_s$, $c_1$ and $c_2$) the master, and slave systems are synchronized. Finally, the proposed controller was implemented in fractional order Duffing-Holmes system. Numerical simulations show the efficiency of the proposed controller in a synchronize task.

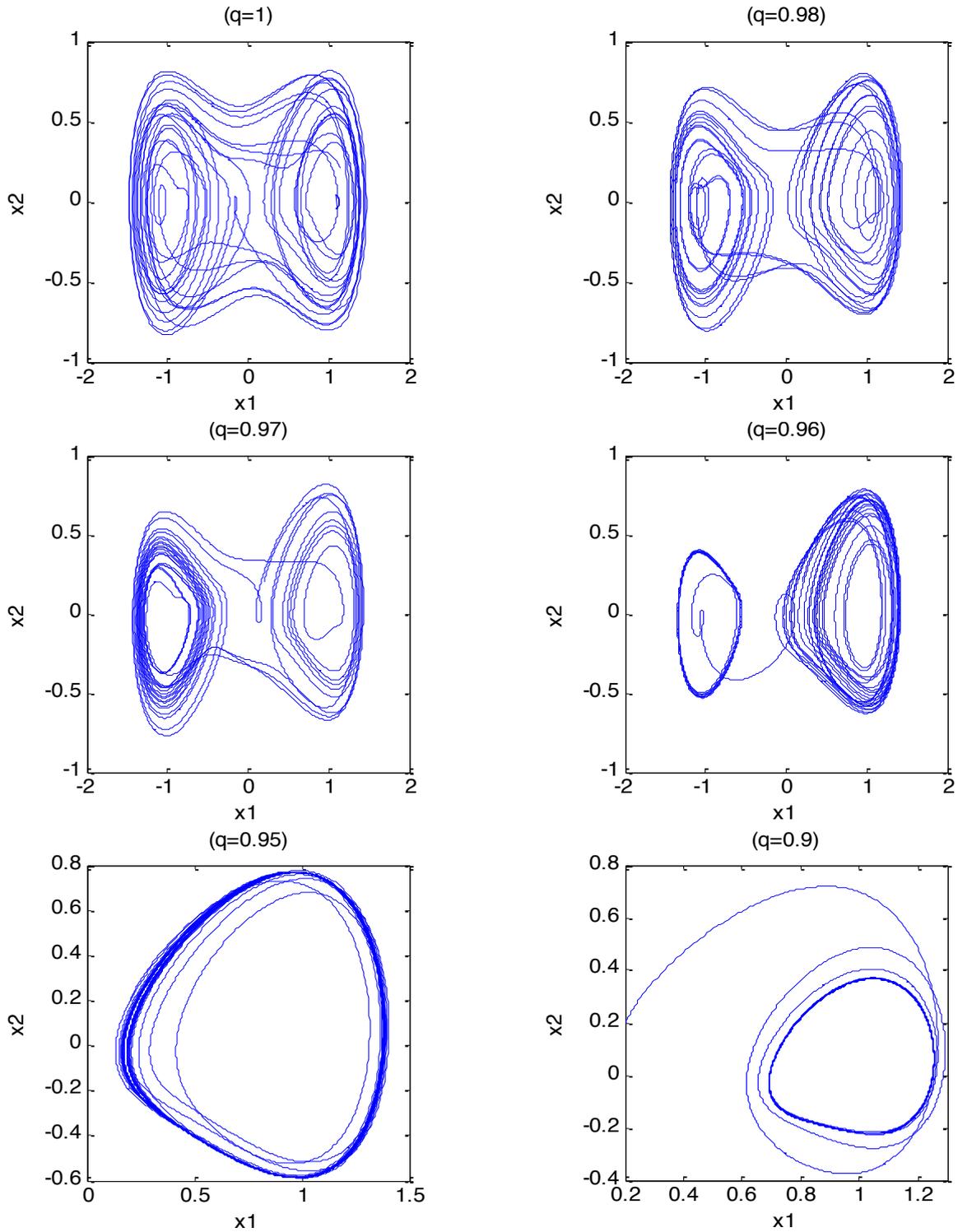

Figure 1: Phase portrait of Duffing-Holmes system vs. different values of the fraction parameter

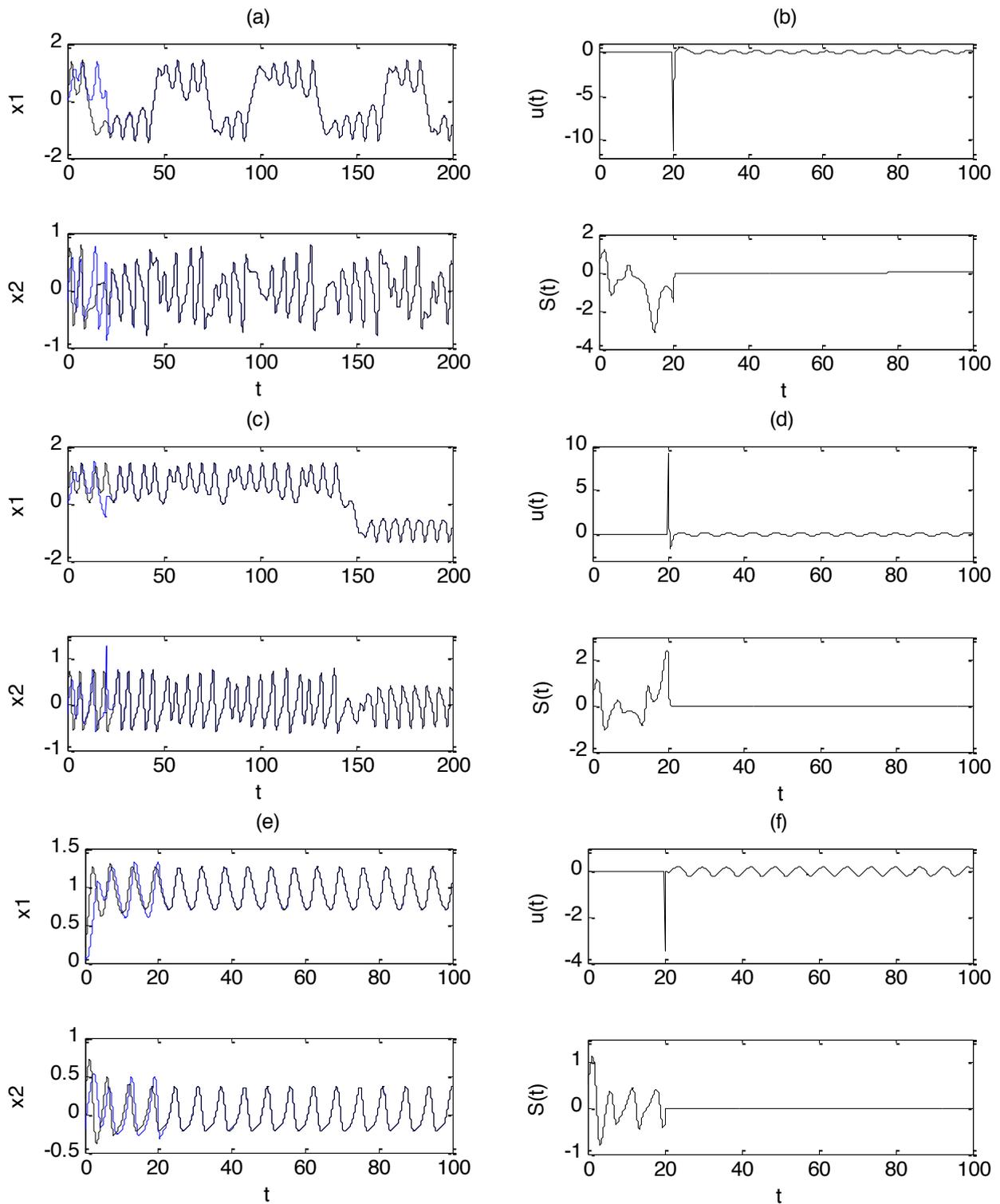

Figure 2: Synchronization results of Fractional order Duffing-Holmes system

REFERENCES:

Ahmad W.M., J.C. Sprott, (2003) , Chaos in fractional-order autonomous nonlinear systems, *Chaos, Solitons Fractals,* 16 339–351.

Arena, P., R. Caponetto, L. Fortuna, D. Porto (1997) , Chaos in a fractional order Duffing system, *in: Proceedings ECCTD*, Budapest, 1259–1262.

Chang W-D and J-J Yan (2005), Adaptive robust PID controller design based on a sliding mode for uncertain chaotic systems,*Chaos, Solitons & Fractals*, **26(1)**, 167-175.

Deng, W., C. Li (2005a), Chaos synchronization of the fractional Lu system, *Physica A*, **353** 61–72.

Deng, W., C. Li (2005b), Synchronization of chaotic fractional Chen system, *J. Phys.Soc. Jpn,* **74 (6)** 1645–1648.

Hartley, T.T., C.F. Lorenzo, H.K. Qammer (1995), Chaos in a fractional order Chua's system, *IEEE Trans. CAS-I*, **42** 485–490.

Ertugrul, M and 0. Kaynak (2000), Neuro-sliding mode control of robotic manipulators, *Mechatronics*, **10**, 239-263.

Li C., G. Chen (2004), Chaos and hyperchaos in the fractional-order Rossler equations, *Physica A: Stat. Mech. Appl.*, **341** 55–61.

Li, C.P W.H. Deng, D. Xu (2006), Chaos synchronization of the Chua system with a fractional order, *Physica A,* **360** 171–185.

Lu, J.G (2005)., Chaotic dynamics and synchronization of fractional-order Arneodo's systems, Chaos, *Solitons Fractals*, **26 (4)** 1125–1133.

Lu, J.G, G. Chen (2006), A note on the fractional-order Chen system, Chaos, Solitons Fractals, **27(3)** 685–688.

Lu, J.G. (2006), Chaotic dynamics of the fractional-order Lu¨ system and its synchronization, Physics Letter A, **354 (4),** 305–311.

Matignon D. **(**1996), Stability results for fractional differential equations with applications to control processing, Computational Engineering in Systems and Application multi-conference, *IMACS, in: IEEE-SMC Proceedings, Lille, France,* **2**, 963–968.

Ott, E., C. Grebogi, J.A. Yorke, (1990), Controlling chaos, Physic Review Letter, **64** 1196–1199.

Phuah, J., J. Lu, and T. Yahagi (2005) , Chattering free sliding mode control in magnetic levitation system, *IEEJ Transaction EIS*, **125 (4)**, 600- 606.

Sheu L.J., H.K. Chen, J.H. Chen, L.M. Tam, W.C. Chen, K.T. Lin, Y. Kang (2008), Chaos in the Newton-Leipnik system with fractional order, *Chaos, Solitons Fractals*, **36,** 98–103.

Slotine, J. J. E. (1991), *Applied Nonlinear Control*, Englewood Cliffs ،New Jersy 07632 ،Prentice Hall.

Slotine, J.E and S. S. Sastry (1983), Tracking control of nonlinear systems using sliding surface with application to robotic manipulators, *International Journal of control,.* **38**, 465 492.

Utkin, V.I. (1977), Variable structure systems with sliding mode, *IEEE Transaction in Automatic Control*, **22,** 212- 222.